\documentclass[prd,twocolumn,showpacs,preprintnumbers,amsmath,amssymb]{revtex4}

\usepackage{graphicx}
\usepackage{dcolumn}
\usepackage{bm}
\usepackage{txfonts}

\begin{document}
\preprint{hep-ph/0504035}

\title{Generalized Boltzmann formalism for oscillating neutrinos}

\author{P. Strack}
 \email{pstrack@physics.arizona.edu}
\affiliation{ Fakult\"at f\"ur Physik und Astronomie, Universit\"at
Heidelberg, 69120 Heidelberg, Germany} \affiliation{Department of
Physics, The University of Arizona, Tucson, Arizona, 85721, USA}

\author{A. Burrows}
 \email{aburrows@as.arizona.edu}
 \affiliation{Steward Observatory and Department of Astronomy, The University of Arizona, Tucson,
 Arizona 85721, USA}

\date{5 April 2005, accepted to Phys. Rev. D}
\begin{abstract}

In the standard approaches to neutrino transport in the simulation
of core-collapse supernovas, one will often start from the classical
Boltzmann equation for the neutrino's spatial, temporal, and
spectral evolution. For each neutrino species, and its
anti-particle, the classical density in phase space, or the
associated specific intensity, will be calculated as a function of
time.  The neutrino radiation is coupled to matter by source and
sink terms on the ``right-hand-side" of the transport equation and
together with the equations of hydrodynamics this set of coupled
partial differential equations for classical densities describes, in
principle, the evolution of core collapse and explosion.  However,
with the possibility of neutrino oscillations between species, a
purely quantum-physical effect, how to generalize this set of
Boltzmann equations for classical quantities to reflect oscillation
physics has not been clear.  To date, the formalisms developed have
retained the character of quantum operator physics involving complex
quantities and have not been suitable for easy incorporation into
standard supernova codes. In this paper, we derive generalized
Boltzmann equations for quasi-classical, real-valued phase-space
densities that retain all the standard oscillation phenomenology,
including the MSW effect (matter-enhanced resonant flavor
conversion), neutrino self-interactions, and the interplay between
decohering matter coupling and flavor oscillations. With this
formalism, any code(s) that can now handle the solution of the
classical Boltzmann or transport equation can easily be generalized
to include neutrino oscillations in a quantum-physically consistent
fashion.


\end{abstract}


\maketitle
\section{Introduction}

Particle oscillations are fundamental for a wide range of
interesting physics: quark mixing by the Cabbibo-Kobayashi-Maskawa
matrix, its leptonic analog for massive neutrinos
\cite{bilenky1,bilenky2}, hypothetical photon-axion and
photon-graviton oscillations in the presence of external magnetic
fields \cite{axion,lowmass}, and $K^{0}-\bar{K}^{0}$ oscillations
\cite{tumunich}. Physically, these quantum systems are coupled to
macroscopic systems and through external interaction their quantum
evolution is altered. One prominent astrophysical context in which
such oscillation for macroscopic systems is important involves the
neutrinos in supernova cores that may execute flavor oscillations
while simultaneously interacting with ambient supernova matter
\cite{burrows3,burrows4,burrows5,raffelt3,pantaleone3,wolfenstein2}.
The primary motivation of the present paper is to provide a
straightforward generalization of the Boltzmann formalism with which
to analyze the kinetics of oscillating neutrinos with collisions. By
taking ensemble-averaged matrix elements of quantum field operators
for mixed particles, following the pioneering work of
\cite{raffelt1,raffelt2,raffelt4,akhiezer,groot}, we obtain
quasi-classical phase-space densities that satisfy real-valued
Boltzmann equations with coupling terms that account for the
neutrino oscillations. The formalism is clear, numerically
tractable, and does not contain operators, complex quantities, or
wavefunctions.

In \S\ref{gbe}, we introduce the Wigner phase-space density operator
approach from which we derive our formalism involving classical
phase-space neutrino flavor densities and their off-diagonal,
overlap correlates.  The latter couple the different flavor states
to account for neutrino oscillations. In \S\ref{simple}, we
demonstrate with several simple examples that the set of equations
reproduces (i) standard flavor oscillations in a vacuum, (ii) flavor
oscillations with absorptive matter coupling (``quantum decoherence"
\cite{raffeltnonabelian}), and (iii) resonant matter-induced flavor
conversion (the MSW effect) for a neutrino beam
\cite{wolfenstein1,bethe,msw}. In \S\ref{conclusions}, we summarize
the salient features of our new approach. We show in Appendix
\ref{appendixnunu} how one might include neutrino and antineutrino
self-interactions into our formalism.

\section{Generalized Boltzmann Equations}
\label{gbe}

To analyze a multi-particle system and reflect its inherent
statistics, one usually sets up the density matrix of the system
\cite{akhiezer,groot}. For classical systems this is the phase-space
density. For quantum fields, the conceptually most similar analog is
the Wigner phase-space density operator \cite{wigner1,wigner2}
\begin{eqnarray}
\rho(\mathbf{r},\mathbf{p},t)=
\int\frac{d^{3}\mathbf{r'}}{(2\pi\hbar)^{3}}e^{-i\mathbf{p}\mathbf{\mathbf{r'}}}
\psi^{\dag}(\mathbf{r}-\frac{1}{2}\mathbf{r'},t)\psi(\mathbf{r}+\frac{1}{2}\mathbf{r'},t)\,\,,\nonumber\\
\label{wignerop}
\end{eqnarray}
where $\psi^{\dag}$ and $\psi$ denote the creation and annihilation
operators, respectively, for the mixing particles of interest
\cite{raffelt1,groot}. $\rho$ works for fermions or bosons
(resulting in different commutation relations of the creation and
annihilation operators) and massive or massless particles. Here, we
average over an ensemble of two neutrino species, for example
$\nu_{e}$'s and $\nu_{\mu}$'s and take the matrix elements in the
number-density basis of Fock space:
\begin{eqnarray} \mathcal{F}=\langle n_{i}|\rho|n_{j}\rangle &=& \left(
      \begin{matrix}
      f_{\nu_{e}}         &  f_{e\mu} \\
      f^{\ast}_{e\mu} &  f_{\nu_{\mu}}  \\
      \end{matrix}
\right)\,\,,\label{matrixelements}
\end{eqnarray}
where the indices $i$, $j$ run over the neutrino flavor, and $^\ast$
means complex conjugation. A generalization to more particle species
is straightforward. The diagonal terms are real-valued and denote
quasi-classical phase-space densities. The off-diagonal entries are
complex-valued macroscopic overlap functions. For completely
decohered ensembles and for non-mixing ensembles the off-diagonal
entries vanish. A Heisenberg-Boltzmann-type equation
\cite{akhiezer,landau,raffelt1,raffeltnonabelian,rudzsky}, which can
heuristically be derived by merging the Boltzmann equation with the
Heisenberg equation, governs the time evolution of the matrix
elements of the Wigner phase-space density operator:
\begin{eqnarray}
\frac{\partial\mathcal{F}}{\partial
t}+\frac{1}{2}\left\{\mathbf{v},\frac{\partial\mathcal{F}}{\partial
\mathbf{r}}\right\}+\frac{1}{2}\left\{\dot{\mathbf{p}},\frac{\partial
\mathcal{F}}{\partial \mathbf{p}}\right\}
=-i\left[\Omega,\mathcal{F}\right]+\mathcal{C}\,\,,\nonumber\\
\label{heisboltz}
\end{eqnarray}
where curly brackets denote matrix anti-commutators, square-brackets
are matrix commutators, $\mathcal{C}$ is the collision matrix and
$\Omega$ is the mixing Hamiltonian. We componentize this equation
for two oscillating neutrino species interacting with a background
medium. Then,
\begin{eqnarray}
\mathcal{C}=\left(\begin{matrix}
\mathcal{C}_{\nu_{e}}& 0\\
0& \mathcal{C}_{\nu_{\mu}}
\end{matrix}
\right)\,\,,
\end{eqnarray}
and the following general mixing Hamiltonian reads:
\begin{eqnarray}
\Omega(\mathbf{r},\mathbf{p},t)&=&\Omega(\varepsilon, \mathbf{r})+
\Omega_{\nu\nu}(\mathbf{r},\mathbf{p},t)-\Omega_{\nu\tilde{\nu}}(\mathbf{r},\mathbf{p},t)\,\,,\nonumber\\
\label{mixhamilton}
\end{eqnarray}
where $\Omega(\varepsilon, \mathbf{r})$ encompasses the vacuum
mixing and the neutrino-matter interaction amplitude for the
$\nu_{e}$ neutrino connected with the matter-induced mass.
$\Omega_{\nu\nu}$ is the off-diagonal mixing contribution from
neutrino-neutrino self-interactions and $\Omega_{\nu\tilde{\nu}}$ is
the analog for neutrino-antineutrino interactions. We show how to
include neutrino self-interactions and neutrino-antineutrino
interaction into our formalism in Appendix \ref{appendixnunu} and
work here with the vacuum and ordinary matter contribution only.
Therefore, we have for $\Omega(\varepsilon, \mathbf{r})$ the
expression
\begin{eqnarray}
\Omega(\varepsilon, \mathbf{r})=\frac{\pi c}{L}\left(\begin{matrix}
-\cos 2 \theta+2A & \sin 2\theta\\
\sin 2\theta & \cos 2 \theta
\end{matrix}
\right)\,\,,\nonumber\\
\end{eqnarray}
where $\theta$ is the neutrino vacuum mixing angle between $\nu_{e}$
and $\nu_{\mu}$ and $A$ is given by \cite{mohapatra,giunti},
\begin{eqnarray}
A=\left(\frac{L}{\pi
c}\right)\frac{2\sqrt{2}G_{F}}{\hbar}n_{e}(\mathbf{r})\,\,,
\end{eqnarray}
where $L$ is the vacuum neutrino oscillation length:
\begin{eqnarray}
L=\frac{4\pi\hbar c\varepsilon}{\Delta m^{2}c^{4}}\,\,.
\end{eqnarray}
$G_{F}$ denotes Fermi's constant, $n_{e}(\mathbf{r})$ the electron
number density, $\varepsilon$ the neutrino energy, $m_{1}$ and
$m_{2}$ are the masses of the neutrino mass eigenstates, and $\Delta
m^{2}=m_{2}^{2}-m_{1}^{2}$. The other variables have their standard
meanings. For antineutrinos the sign of $A$ is reversed. Note that
$A$, through $n_e$, is in general a function of spatial position. In
this paper, we do not take into account the effects of microscopic
density fluctuations on the mixing in the ensemble \cite{sawyer2}.
In other words, we assume that the scale of the spatial variation of
the phase-space density is larger than the neutrino de Broglie
wavelength. Similarly, we assume small external forces. With these
reasonable assumptions, we can ignore the off-diagonal terms on the
left-hand-side of Eq. (\ref{heisboltz}). Defining the real part of
the off-diagonal macroscopic overlap in Eq. (\ref{matrixelements})
as
\begin{eqnarray}
f_{r}=\frac{1}{2}\left(f_{e\mu}+f^{\ast}_{e\mu}\right)\,\,,
\label{offdiare}
\end{eqnarray}
and the corresponding imaginary part as
\begin{eqnarray}
f_{i}=\frac{1}{2i}\left(f_{e\mu}-f^{\ast}_{e\mu}\right)\,\,,
\label{offdiaim}
\end{eqnarray}
respectively, we find the generalized Boltzmann equations for
real-valued quantities:
\begin{eqnarray}
\frac{\partial f_{\nu_{e}}}{\partial
t}+\mathbf{v}\cdot\frac{\partial f_{\nu_{e}}}{\partial\mathbf{r}}+
\mathbf{\dot{p}}\cdot\frac{\partial
f_{\nu_{e}}}{\partial\mathbf{p}}&=&
-\frac{2\pi c}{L}f_{i}\sin2\theta +\mathcal{C}_{\nu_{e}}\nonumber\\
\frac{\partial f_{\nu_{\mu}}}{\partial
t}+\mathbf{v}\cdot\frac{\partial f_{\nu_{\mu}}}{\partial\mathbf{r}}+
\mathbf{\dot{p}}\cdot\frac{\partial
f_{\nu_{\mu}}}{\partial\mathbf{p}}&=&
\frac{2\pi c}{L}f_{i}\sin2\theta +\mathcal{C}_{\nu_{\mu}}\nonumber\\
\frac{\partial f_{r}}{\partial t}+\mathbf{v}\cdot\frac{\partial
f_{r}}{\partial \mathbf{r}}+\mathbf{\dot{p}} \cdot\frac{\partial
f_{r}}{\partial \mathbf{p}}&=&
-\frac{2\pi c}{L}f_{i}\left(\cos2\theta-A\right) \nonumber\\
\frac{\partial f_{i}}{\partial t}+\mathbf{v}\cdot\frac{\partial
f_{i}}{\partial \mathbf{r}}+\mathbf{\dot{p}} \cdot\frac{\partial
f_{i}}{\partial \mathbf{p}}&=&\frac{2\pi
c}{L}\Big(\frac{f_{\nu_{e}}-f_{\nu_{\mu}}}{2}\sin2\theta+\nonumber\\
&&f_{r}\left(\cos2\theta-A\right)\Big)\,\,.\nonumber\\
\label{genericboltz}
\end{eqnarray}
The neutrino oscillations are incorporated with new sink and source
terms indirectly coupling the standard Boltzmann equations for
$f_{\nu_{e}}$ and $f_{\nu_{\mu}}$ through the off-diagonal
macroscopic overlap functions $f_{r}$ and $f_{i}$. The number of
equations is increased from two to four. In principle, one can
insert blocking factors for both the collision terms and the
oscillation terms. Note that in the absence of collisions,
Liouville's theorem for the total lepton specific intensity,
$d/dt\left(f_{\nu_{e}}+f_{\nu_{\mu}}\right)=0$, is recovered.

It is instructive to rewrite Eq. (\ref{genericboltz}) in terms of
the specific intensities of the neutrino radiation field. We use the
one-to-one relation \cite{burrows1, mihalas} between the invariant
phase-space densities, $f_{\nu_{e}}$, $f_{\nu_{\mu}}$ and the
specific intensities, $I_{\nu_{e}}$, $I_{\nu_{\mu}}$, and define
accordingly:
\begin{eqnarray}
I_{\nu_{e}}=\frac{\varepsilon^{3}f_{\nu_{e}}}{(2\pi
\hbar)^{3}c^{2}}\,\,,\quad\quad
I_{\nu_{\mu}}=\frac{\varepsilon^{3}f_{\nu_{\mu}}}{(2\pi
\hbar)^{3}c^{2}}\,\,.
\end{eqnarray}
For the off-diagonal macroscopic overlap analogs one defines
\begin{eqnarray}
\mathcal{R}_{e\mu}=\frac{\varepsilon^{3}f_{r}}{(2\pi
\hbar)^{3}c^{2}}\,\,,\quad\quad
\mathcal{I}_{e\mu}=\frac{\varepsilon^{3}f_{i}}{(2\pi
\hbar)^{3}c^{2}}\,\,. \label{reps}
\end{eqnarray}
The generalized Boltzmann equations in the laboratory (Eulerian
frame) for the radiation field of two oscillating neutrino species
with collisions are then:
\begin{widetext}
\begin{eqnarray}
\frac{1}{c}\frac{\partial I_{\nu_{e}}}{\partial
t}+\frac{\mathbf{v}}{c}\cdot\frac{\partial
I_{\nu_{e}}}{\partial\mathbf{r}}
+\frac{\varepsilon^{3}\mathbf{\dot{p}}}{c}\cdot\frac{\partial\left(
I_{\nu_{e}}\varepsilon^{-3}\right)}{\partial \mathbf{p}}&=&
-\frac{2\pi}{L}\mathcal{I}_{e\mu}\sin2\theta+\mathcal{C}'_{\nu_{e}}\nonumber\\
\frac{1}{c}\frac{\partial I_{\nu_{\mu}}}{\partial
t}+\frac{\mathbf{v}}{c}\cdot\frac{\partial
I_{\nu_{\mu}}}{\partial\mathbf{r}}
+\frac{\varepsilon^{3}\mathbf{\dot{p}}}{c}\cdot\frac{\partial\left(
I_{\nu_{\mu}}\varepsilon^{-3}\right)}{\partial \mathbf{p}}&=&
\frac{2\pi}{L}\mathcal{I}_{e\mu}\sin2\theta+
\mathcal{C}'_{\nu_{\mu}}\nonumber\\
\frac{1}{c}\frac{\partial \mathcal{R}_{e\mu}}{\partial
t}+\frac{\mathbf{v}}{c}\cdot\frac{\partial
\mathcal{R}_{e\mu}}{\partial\mathbf{r}}+\frac{\varepsilon^{3}\mathbf{\dot{p}}}{c}\cdot\frac{\partial\left(
\mathcal{R}_{e\mu}\varepsilon^{-3}\right)}{\partial \mathbf{p}}&=&
-\frac{2\pi}{L}\left(\cos2\theta-A\right)\mathcal{I}_{e\mu}\nonumber\\
\frac{1}{c}\frac{\partial \mathcal{I}_{e\mu}}{\partial
t}+\frac{\mathbf{v}}{c}\cdot\frac{\partial
\mathcal{I}_{e\mu}}{\partial\mathbf{r}}
+\frac{\varepsilon^{3}\mathbf{\dot{p}}}{c}\cdot\frac{\partial\left(
\mathcal{I}_{e\mu}\varepsilon^{-3}\right)}{\partial
\mathbf{p}}&=&\frac{2\pi}{L}\left(
\frac{I_{\nu_{e}}-I_{\nu_{\mu}}}{2}\sin2\theta
+\left(\cos2\theta-A\right)\mathcal{R}_{e\mu}\right)\,\,.
\label{radiationfield}
\end{eqnarray}
\end{widetext}
These equations and Eq. (\ref{genericboltz}) are our major results.
The collision terms can be conveniently written as \cite{burrows1}
\begin{eqnarray}
\mathcal{C}'_{\nu_{e}}&=&-\kappa^{s}_{\nu_{e}}I_{\nu_{e}}+\kappa^{a
}_{\nu_{e}}\left(\frac{B_{\nu_{e}}-I_{\nu_{e}}}{1-\mathcal{F}^{eq}_{\nu_{e}}}\right)\nonumber\\
&&+\frac{\kappa^{s}_{\nu_{e}}}{4\pi}
\int\Phi_{\nu_{e}}(\mathbf{\Omega},\mathbf{\Omega'})I_{\nu_{e}}(\mathbf{\Omega'})d\Omega'\,\,,
\end{eqnarray}
where $\Phi_{\nu_{e}}$ is a phase function for scattering into the
beam integrated over the solid angle $d\Omega'$, $\kappa^{a
}_{\nu_{e}}$ is the sum of all absorption processes
$\sum_{i}n_{i}\sigma^{a}_{i}$, where $n_{i}$ is the number density
of matter species $i$ and $\sigma^{a}_{i}$ denotes the absorption
cross sections (for scattering processes the superscript $a$ is
replaced with $s$), $\mathcal{F}^{eq}_{\nu_{e}}$ is the equilibrium
Fermi-Dirac occupation probability, and $B_{\nu_{e}}$ is the
corresponding blackbody specific intensity. Changing the subscript
from $\nu_{e}$ to $\nu_{\mu}$ yields the corresponding parameters
for the $\nu_{\mu}$'s. For sterile neutrinos, one substitutes
$\nu_{s}$ for $\nu_{\mu}$ and sets the scattering and absorption
terms to zero. Similarly, one can write a set of equations for
antineutrinos with different collision terms and with the sign of
$A$ reversed. Neutrino and antineutrino evolution are implicitly
coupled through pair processes. If self-interactions are included as
we show in Appendix \ref{appendixnunu}, neutrino and antineutrino
evolution are nonlinearly coupled.

\section{\mbox{Simple Tests of the New Formalism}}
\label{simple}

\subsection{Oscillations with absorptive matter coupling}
\label{sterilesec}

It is straightforward to show that our set of generalized coupled
Boltzmann equations behaves as we would expect from our experience
with the standard wave function approach. As our first example, we
solve Eq. (\ref{radiationfield}) for $\nu_{e}$ -- $\nu_{\mu}$
oscillations in box of isotropic neutrinos that can also experience
decohering absorption on matter. Note that in reality neutrino
interactions only play a role at densities where observable neutrino
oscillations are suppressed \cite{wolfenstein2}. To demonstrate the
expected limiting behavior of our formalism we artificially ``turn
off'' matter suppression by setting the matter term $A$ to zero.
This is not the situation found in nature. We define the approximate
oscillation time
\begin{eqnarray}
t_{osc}&\simeq&\frac{L}{2\pi c}=\frac{2\hbar\varepsilon}{\Delta
m^{2}c^{4}}\,\,,
\end{eqnarray}
and set the interaction rate of the $\nu_{e}$'s equal to the
interaction rate of the $\nu_{\mu}$'s to define the characteristic
absorption time
\begin{eqnarray}
t^{\nu_{e}}_{col}=\frac{1}{\kappa^{a
\ast}_{\nu_{e}}c}=\frac{\left(1-\mathcal{F}^{eq}_{\nu_{e}}\right)}{cN_{A}\rho
Y_{n}\sigma^a_{\nu_{e}n}}\,\,.
\end{eqnarray}
$N_{A}$ denotes Avogadro's number, $Y_{n}$ is the neutron fraction
per nucleon, and $\sigma^a_{\nu_{e}n}$ is the cross section for
absorption on neutrons (see Appendix \ref{appendixcross}). We define
the ratio of the oscillation to the absorption timescale:
\begin{eqnarray}
\alpha=\frac{t_{osc}}{t^{\nu_{e}}_{col}}\,\,,
\end{eqnarray}
and the dimensionless time coordinate $\tau$:
\begin{eqnarray}
\tau=\frac{t}{t_{osc}}\,\,.
\end{eqnarray}
Furthermore, we set $B_{\nu_{e}}=B_{\nu_{\mu}}$ and denote the
dimensionless specific intensities and the off-diagonal macroscopic
overlap functions that are normalized to the blackbody intensity
with a hat. The resulting dimensionless version of Eq.
(\ref{radiationfield}) reads:
\begin{eqnarray}
\frac{\partial \widehat{I}_{\nu_{e}}}{\partial
\tau}&=&-\widehat{\mathcal{I}}_{e\mu}\sin2\theta+\alpha\left(1-\widehat{I}_{\nu_{e}}\right)\nonumber\\
\frac{\partial \widehat{I}_{\nu_{\mu}}}{\partial
\tau}&=&\widehat{\mathcal{I}}_{e\mu}\sin2\theta+\alpha\left(1-\widehat{I}_{\nu_{\mu}}\right)\nonumber\\
\frac{\partial \widehat{\mathcal{R}}_{e\mu}}{\partial
\tau}&=&-\widehat{\mathcal{I}}_{e\mu}\cos2\theta \nonumber\\
\frac{\partial \widehat{\mathcal{I}}_{e\mu}}{\partial
\tau}&=&\frac{\widehat{I}_{\nu_{e}}-\widehat{I}_{\nu_{\mu}}}{2}\sin2\theta+
\widehat{\mathcal{R}}_{e\mu}\cos2\theta\,\,,\nonumber\\
\label{dimlessneutrons}
\end{eqnarray}
where the scattering in and out of the beam has been canceled due to
the assumption of isotropy. In a vacuum ($\alpha=0$) and with the
initial conditions $\widehat{I}_{\nu_{e}}=1$,
$\widehat{I}_{\nu_{\mu}}=0$, and consequently no off-diagonal
overlap terms at $\tau=0$, we obtain
\begin{eqnarray}
\widehat{I}_{\nu_{e}}(\tau)&=&1-\sin^{2}2\theta\sin^{2}\left(\frac{\tau}{2}\right)\nonumber\\
\widehat{I}_{\nu_{\mu}}(\tau)&=&\sin^{2}2\theta\sin^{2}\left(\frac{\tau}{2}\right)\nonumber\\
\widehat{\mathcal{R}}_{e\mu}(\tau)&=&\frac{1}{2}\sin2\theta\cos2\theta\left(\cos\tau-1\right)\nonumber\\
\widehat{\mathcal{I}}_{e\mu}(\tau)&=&\frac{1}{2}\sin2\theta\sin\tau\,\,.
\label{homo}
\end{eqnarray}
This behavior of the radiation field is unambiguously identical to
the probability density obtained by squaring the amplitude of a
single-neutrino wave function in a beam. The off-diagonal terms
representing the macroscopic overlap peak when mixing of $\nu_{e}$
and $\nu_{\mu}$ neutrinos is maximal and vanish when the ensemble is
single-flavored. In matter and for the initial conditions
$\widehat{I}_{\nu_{\mu}}=\widehat{\mathcal{R}}_{e\mu}=\widehat{\mathcal{I}}_{e\mu}=0$
and $\widehat{I}_{\nu_{e}}\neq0$, one can derive an harmonic
oscillator equation for the early rate of evolution of
$\widehat{I}_{\nu_{e}}$:
\begin{eqnarray}
\frac{\partial^{2}\widehat{I}_{\nu_{e}}}{\partial
\tau^{2}}+\left(\frac{1}{2}-\alpha^{2}\right)\widehat{I}_{\nu_{e}}=\text{const}\,\,.
\end{eqnarray}
As expected, for $\alpha\ll1$, the early time dependence of the
solution is predominantly sinusoidal: neutrino oscillations
dominate. For $\alpha\gg1$, an exponential decay/increase dominates.
The timescale then is $1/\alpha$.

\begin{figure}
\includegraphics[width=82mm]{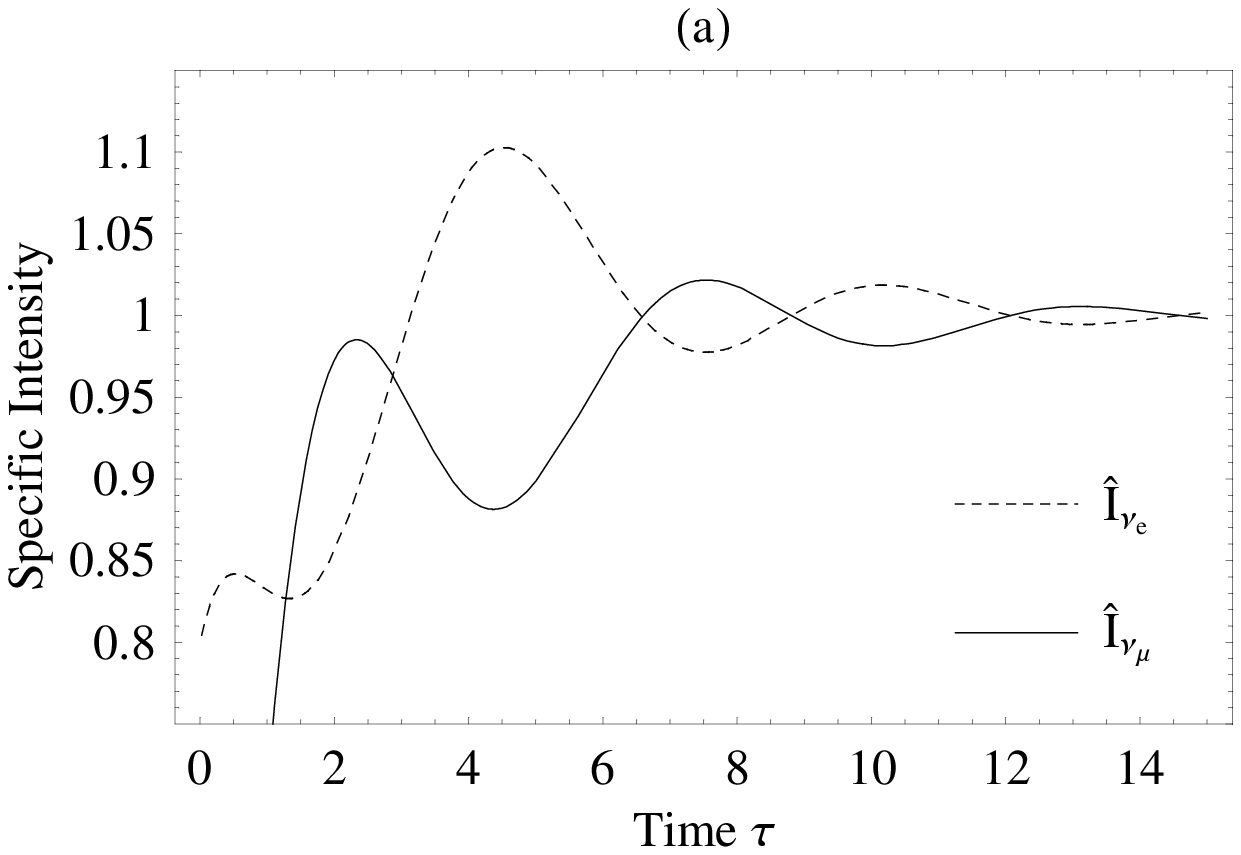}\\[3mm]
\includegraphics[width=82mm]{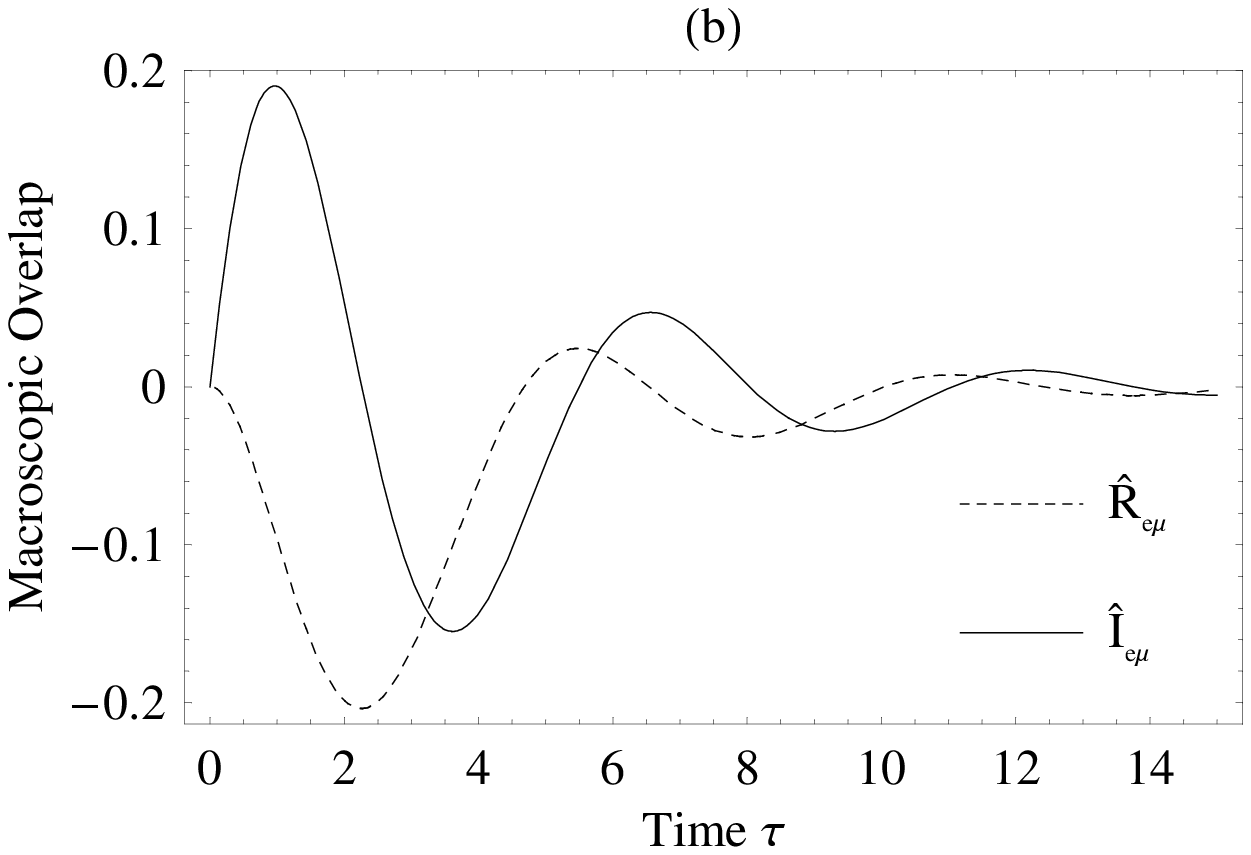}
\caption{$\nu_{e}$ -- $\nu_{\mu}$ oscillations of isotropic
neutrinos in a box with absorptive matter coupling. (\textbf{a})
Specific intensities. (\textbf{b}) Off-diagonal macroscopic overlap
functions. Parameters:
$\varepsilon_{\nu_{e}}=\varepsilon_{\nu_{\mu}}=10$ MeV,
$\rho=8\times 10^{12}\,\text{g cm}^{-3}$, T $=5$ MeV, and from the
large-mixing-angle solution (LMA)\cite{superk}: $\sin2\theta=0.9$,
and $\Delta m^{2}=6.9\times10^{-5}$ eV$^{2}$ . $A$ is artificially
set to zero.}\label{box}
\end{figure}

In Fig. \ref{box}, we depict the solutions to Eq.
(\ref{dimlessneutrons}) for oscillating $\nu_{e}$'s in a box with
nucleons. Initial conditions are
$\widehat{I}_{\nu_{\mu}}=\widehat{\mathcal{R}}_{e\mu}=\widehat{\mathcal{I}}_{e\mu}=0$
and $\widehat{I}_{\nu_{e}}=0.8$ . For this example, flavor
oscillations and collisions happen on the same timescale. The
ensemble is guided to flavor and radiative equilibrium. Coherent
flavor oscillations are disrupted by absorptive collisions.
Asymptotically, the diagonal specific intensities for the
$\nu_{e}$'s and $\nu_{\mu}$'s equilibrate. Absorption on neutrons,
and by detailed balance, the resulting emissivity, drive the
$\nu_{e}$'s to the blackbody intensity. The oscillation amplitude
decreases with time; the quantum evolution of the system is
decohered through absorptive coupling with matter (``quantum
decoherence" \cite{raffeltnonabelian}). The real part of the
off-diagonal overlap, $\widehat{\mathcal{R}}_{e\mu}$, takes
predominantly negative values whereas the imaginary part,
$\widehat{\mathcal{I}}_{e\mu}$, oscillates symmetrically around
zero. Both vanish asymptotically and no oscillations persist.
\subsection{Matter-enhanced resonant flavor conversion}
\label{matter} To demonstrate that our formalism contains the MSW
effect \cite{wolfenstein1,bethe,msw}, we solve Eq.
(\ref{radiationfield}) for a monoenergetic one-dimensional neutrino
beam propagating down a density profile for which resonant
matter-enhanced flavor conversion takes place. We define the
dimensionless distance coordinate in terms of the oscillation
length:
\begin{eqnarray}
\widehat{x}=x\,\frac{2\pi}{L}\,\,,
\end{eqnarray}
and the dimensionless matter-induced mass term in terms of its
resonance value:
\begin{eqnarray}
\widehat{A}=\frac{A}{A_{res}}=\frac{A}{\cos2\theta}\,\,.
\label{dimlessvariables2}
\end{eqnarray}
The beam passes the resonance density for $\widehat{A}=1$. We
analyze the following dimensionless version of Eq.
(\ref{radiationfield}):
\begin{eqnarray}
\frac{\partial I_{\nu_{e}}}{\partial\widehat{x}}&=&-\mathcal{I}_{e\mu}\sin2\theta\nonumber\\
\frac{\partial I_{\nu_{\mu}}}{\partial \widehat{x}}&=&\mathcal{I}_{e\mu}\sin2\theta\nonumber\\
\frac{\partial\mathcal{R}_{e\mu
}}{\partial \widehat{x}}&=&-\cos2\theta\left(1-\widehat{A}\right)\mathcal{I}_{e\mu}\nonumber\\
\frac{\partial\mathcal{I}_{e\mu}}{\partial
\widehat{x}}&=&\frac{I_{\nu_{e}}-I_{\nu_{\mu}}}
{2}\sin2\theta+\cos2\theta\left(1-\widehat{A}\right)\mathcal{R}_{e\mu}\,\,,\nonumber\\
\label{eq:msw}
\end{eqnarray}
where $\mathcal{C}'_{\nu_{e}}$ and $\mathcal{C}'_{\nu_{\mu}}$ have
been set to zero.
\begin{figure}
\includegraphics[width=82mm]{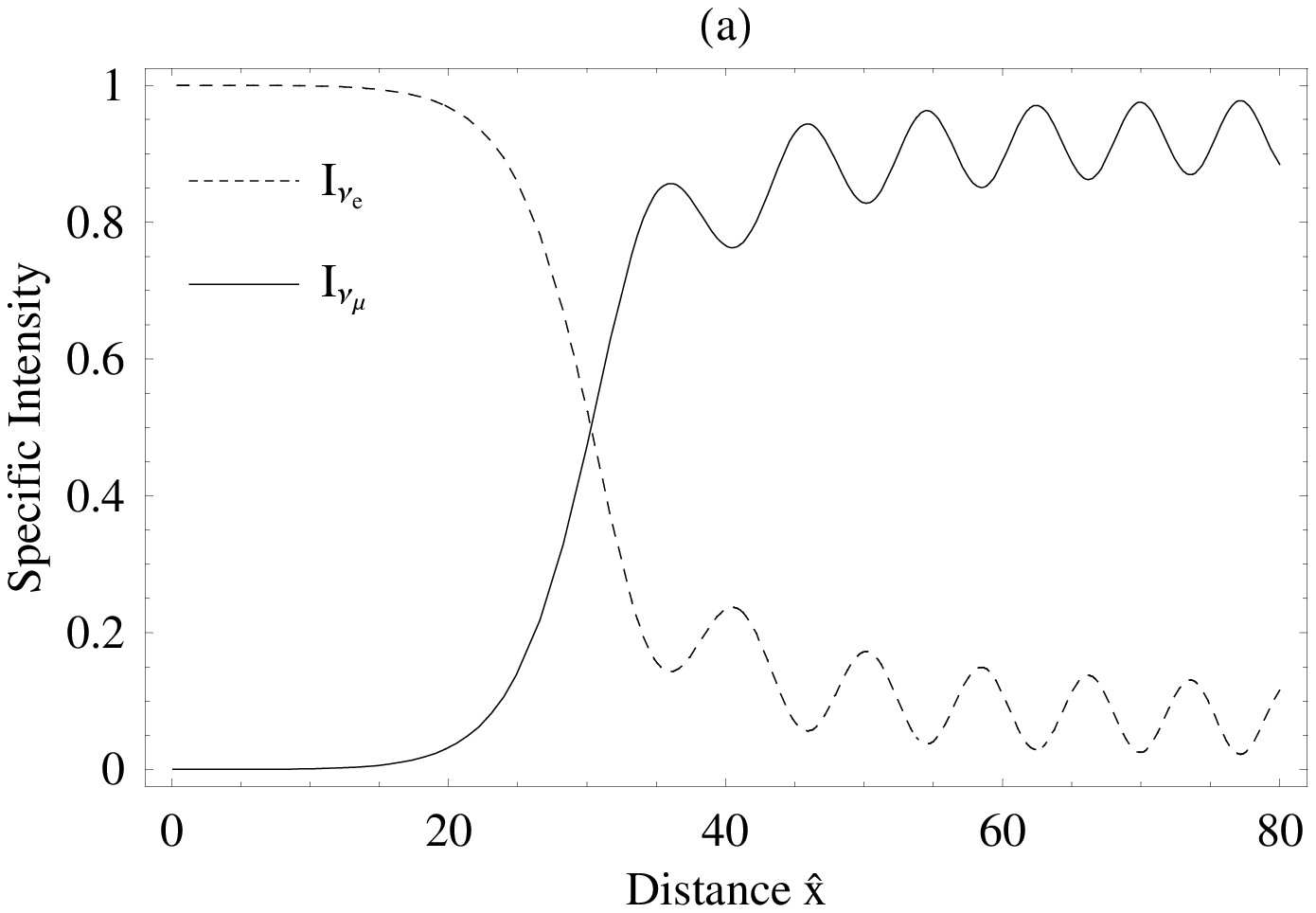}\\[3mm]
\includegraphics[width=82mm]{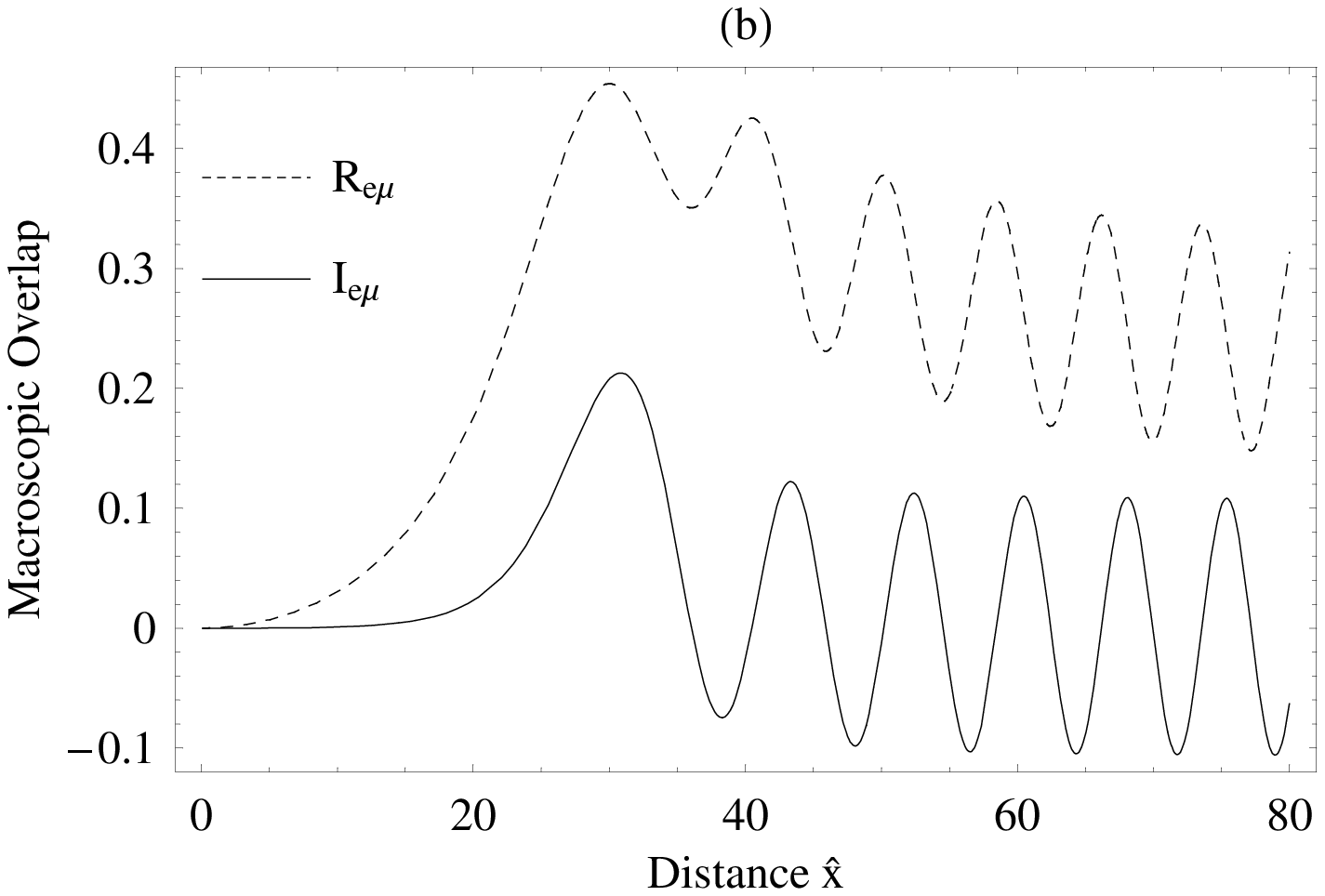}
\caption{MSW effect. (\textbf{a}) Specific intensities. (\textbf{b})
Off-diagonal macroscopic overlap functions. An initially $\nu_{e}$
beam propagates down the density profile
$\frac{\beta}{\widehat{x}^{2}}$, where $\beta=900$.} \label{msw}
\end{figure}

In Fig. \ref{msw}, we depict the solutions to Eq. (\ref{eq:msw}) for
a density profile of $A=\frac{\beta}{\widehat{x}^{2}}$. We set
$\beta=900$ and thus ensure that the scale of spatial
inhomogeneities is large compared to the microscopic length scales
such as the neutrino de Broglie wavelength and the oscillation
length. Initially, the beam contains only $\nu_{e}$ neutrinos. The
mixing angle is arbitrarily taken to be $\sin^{2}2\theta=0.18$ (at
present the LMA is favored \cite{superk}). From Fig. \ref{msw}, it
is clear that at the resonance density,
$\widehat{x}=\sqrt{\beta}=30$, the flavor composition of the beam is
radically altered. For higher values of $\widehat{x}$, the beam
executes vacuum oscillations. In this illustrative problem the
density at production is much greater than the resonance density.
Then, the spatially averaged survival probability of a $\nu_{e}$
neutrino going from matter to free space should be \cite{parke}
\begin{eqnarray}
\langle
P(\nu_{e}\rightarrow\nu_{e})\rangle\approx\left(1-P_{x}\right)\sin^{2}\theta+P_{x}\cos^{2}\theta\,\,,
\label{survival}
\end{eqnarray}
where $P_{x}$ is the Landau-Zener probability for non-adiabatic
transitions. For the chosen density profile, propagation is
adiabatic and $P_{x}=0$. The averaged $\nu_{e}$ survival probability
in Fig. \ref{msw} converges toward $\sim 0.05$, which is congruent
with the value predicted in Eq. (\ref{survival}):
$\sin^{2}\theta\sim0.05$. For high densities ($\widehat{x}\leq20$),
matter suppression is severe. No flavor oscillations happen and the
off-diagonal overlap functions $\mathcal{R}_{e\mu}$ and
$\mathcal{I}_{e\mu}$ are close to zero. In free space for
$\widehat{x}>60$, the imaginary part $\mathcal{I}_{e\mu}$ oscillates
symmetrically around zero whereas the real part $\mathcal{R}_{e\mu}$
is positive.

The solution given in Fig. (\ref{msw}) is numerically equivalent to
the solution obtained using the standard wavefunction formalism
\cite{giunti,wolfenstein1,bethe,msw}:
\begin{eqnarray}
i\frac{\partial}{\partial \widehat{x}}\left(\begin{matrix}
\psi_{\nu_{e}}\\
\psi_{\nu_{\mu}}\end{matrix}\right)=\frac{1}{2}\left(\begin{matrix}
-\cos2\theta\left(1-2\widehat{A}\right)&
\sin2\theta\\
\sin2\theta&\cos2\theta\end{matrix}\right) \left(\begin{matrix}
\psi_{\nu_{e}}\\
\psi_{\nu_{\mu}}\end{matrix}\right)\,\,,\nonumber\\
 \label{schroed}
\end{eqnarray}
when we identify the specific intensities with the probability
densities
\begin{eqnarray}
I_{\nu_{e}}&\leftrightarrow&|\psi_{\nu_{e}}|^{2}\nonumber\\
I_{\nu_{\mu}}&\leftrightarrow&|\psi_{\nu_{\mu}}|^{2}\,\,,
\label{match1}
\end{eqnarray}
and the macroscopic overlap functions with
\begin{eqnarray}
\mathcal{R}_{e\mu}&\leftrightarrow&\frac{1}{2}\left(\psi_{\nu_{e}}\psi^{\ast}_{\nu_{\mu}}
+\psi^{\ast}_{\nu_{e}}\psi_{\nu_{\mu}}\right)\nonumber\\
\mathcal{I}_{e\mu}&\leftrightarrow&\frac{1}{2i}\left(\psi_{\nu_{e}}\psi^{\ast}_{\nu_{\mu}}
-\psi^{\ast}_{\nu_{e}}\psi_{\nu_{\mu}}\right)\,\,. \label{match2}
\end{eqnarray}
Thus, our Boltzmann formalism is completely consistent with the
existing description.

\section{Conclusions}
\label{conclusions} In this paper, we have derived a generalized set
of Boltzmann equations for real-valued phase-space densities of
oscillating neutrinos interacting with a background medium. The
off-diagonal functions of the Wigner phase-space density matrix
representing macroscopic overlap are explicitly included and serve
to couple the flavor states to reflect neutrino oscillation physics.
Conceptually, we have reduced the time evolution of creation and
annihilation operators to that of real valued phase-space densities
without losing quantum-physical accuracy. Important quantum effects
such as matter-enhanced resonant flavor conversion and
``decoherence"\cite{raffeltnonabelian} through matter coupling are
correctly incorporated. The generalized Boltzmann equations are
simple and very similar to the equations of classical transport
theory. Neutrino oscillations are incorporated by new sink and
source terms that indirectly couple the expanded set of equations.
We have shown how to include neutrino self-interactions in our
formalism. The self-interactions nonlinearly couple neutrino and
antineutrino evolution and therefore for the most generic case one
has to deal with eight non-trivially coupled equations for a
two-flavor ensemble and their antiparticles.

Using this formalism, codes that now solve the standard Boltzmann
equations for the classical neutrino phase-space density
($f_{\nu_i}$), or which address its angular and/or energy moments,
can straightforwardly be reconfigured by the simple addition of
source terms and similar transport equations for overlap densities
that have the same units as $f_{\nu_i}$, to incorporate neutrino
oscillations in a quantum-physically consistent fashion.

\begin{acknowledgments}
P.S. thanks the German Fulbright-Commission, and the
Baden-W\"urttemberg Stiftung for support, and Steward Observatory
for kind hospitality. We acknowledge support for this work through
the SciDAC program of the Department of Energy under grant number
DE-FC02-01ER41184. We thank Zackaria Chacko for helpful
conversations and for reading the manuscript. We also thank Casey
Meakin and Martin C. Prescher for useful discussions.
\end{acknowledgments}

\appendix
\section{Self-Interactions}
\label{appendixnunu}
Evolution in a neutrino background is
nontrivial \cite{fuller,pantaleone1,pantaleone2,pantaleone4}. To
address this, one can retain the one-particle character of
description and neglect any effects arising from coherent many-body
state formation or flavor entanglement
\cite{friedland,cecilia,bell,sawyer2004,sawyer2005}. The local
low-energy four Fermi-interaction Hamiltonian of neutrino-neutrino
scattering is given by
\begin{eqnarray}
\Omega^{\mathbf{pq}}_{\nu\nu}=\frac{G_{F}}{\sqrt{2}}\left(\sum_{i}
\bar{\psi}^{i}_{\mathbf{q}}\gamma^{\mu}\psi^{i}_{\mathbf{q}}\right)
\left(\sum_{j}\bar{\psi}^{j}_{\mathbf{q}}\gamma_{\mu}\psi^{j}_{\mathbf{q}}\right)\,\,,
\end{eqnarray}
where the sum is over all neutrino flavors. This effective
Hamiltonian must satisfy U(N) flavor symmetry for a system of N
flavors \cite{pantaleone1}. For a two-flavor system consisting of
$\nu_{e}$'s and $\nu_{\mu}$'s one can rewrite this Hamiltonian to
accentuate its off-diagonal character \cite{pantaleone2,friedland}:
\begin{eqnarray}
\Omega^{\mathbf{pq}}_{\nu\nu}&=&\beta\left(1-\cos\theta^{\mathbf{pq}}\right)\times\nonumber\\
&&\left[\Big|\psi_{\nu_{e}}^{\mathbf{q}}\Big|^{2}
+\Big|\psi_{\nu_{\mu}}^{\mathbf{q}}\Big|^{2}+ \left(\begin{matrix}
\Big|\psi_{\nu_{e}}^{\mathbf{q}}\Big|^{2} & \psi_{\nu_{e}}^{\mathbf{q}}\psi_{\nu_{\mu}}^{\mathbf{q}\ast}\\
\psi_{\nu_{e}}^{\mathbf{q}\ast}\psi_{\nu_{\mu}}^{\mathbf{q}}
&\left|\psi_{\nu_{\mu}}^{\mathbf{q}}\right|^{2}\end{matrix}
\right)\right]\,\,,\nonumber\\
\label{nunu}
\end{eqnarray}
where the coupling coefficient includes the angle between the ``test
neutrino'' with momentum $\mathbf{p}$ and the background neutrino
with momentum $\mathbf{q}$ and the coupling strength is given by
$\beta=\frac{\sqrt{2}G_{F}}{\hbar}$. The neutrino fields of the
background neutrino with momentum $\mathbf{q}$ are normalized such
that
\begin{eqnarray}
\int dV
\left(\Big|\psi_{\nu_{e}}^{\mathbf{q}}\Big|^{2}+\Big|\psi_{\nu_{\mu}}^{\mathbf{q}}\Big|^{2}\right)=1\,\,.
\end{eqnarray}
We now match the above expressions and convert them into our
quasi-classical phase-space densities and the macroscopic overlap
functions. The $\nu$--$\nu$ mixing Hamiltonian for test neutrinos
with momentum $\mathbf{p}$ in ensemble-averaged form is denoted by
\begin{eqnarray}
\Omega_{\nu\nu}(\mathbf{p},\mathbf{r},t)=\left(\begin{matrix}
B_{\nu_{e}}  & B_{r}+iB_{i}\\
B_{r}-iB_{i} & B_{\nu_{\mu}}\end{matrix} \right)\,\,,
\end{eqnarray}
where, while throwing away the overall phase term in Eq.
(\ref{nunu}) proportional to the identity matrix, we have for the
diagonal elements:
\begin{eqnarray}
B_{\nu_{e}}(\mathbf{p},\mathbf{r},t)&=&\beta\int
d^{3}\mathbf{q}\left(1-\cos\theta^{\mathbf{pq}}\right)f_{\nu_{e}}(\mathbf{q},\mathbf{r},t)\nonumber\\
B_{\nu_{\mu}}(\mathbf{p},\mathbf{r},t)&=&\beta\int
d^{3}\mathbf{q}\left(1-\cos\theta^{\mathbf{pq}}\right)f_{\nu_{\mu}}(\mathbf{q},\mathbf{r},t)\,\,.\nonumber\\
\end{eqnarray}
We used the ``matching'' from our formalism to the wavefunction
formalism as prescribed in Eqs. (\ref{match1}) and (\ref{match2}).
The momentum integration goes over all momenta $\mathbf{q}$ in the
ensemble. For the off-diagonal elements we write in our notation,
\begin{eqnarray} B_{r}(\mathbf{p},\mathbf{r},t)&=&\beta\int
d^{3}\mathbf{q}\left(1-\cos\theta^{\mathbf{pq}}\right)f_{r}(\mathbf{q},\mathbf{r},t)\nonumber\\
B_{i}(\mathbf{p},\mathbf{r},t)&=&\beta\int
d^{3}\mathbf{q}\left(1-\cos\theta^{\mathbf{pq}}\right)f_{i}(\mathbf{q},\mathbf{r},t)\,\,,\nonumber\\
\end{eqnarray}
where we have used the variables of Eqs. (\ref{offdiare}) and
(\ref{offdiaim}). For the neutrino-antineutrino interaction there is
\begin{eqnarray}
\Omega_{\nu\tilde{\nu}}(\mathbf{p},\mathbf{r},t)=\left(\begin{matrix}
\tilde{B}_{\nu_{e}} & \tilde{B}_{r}+i\tilde{B}_{i}\\
\tilde{B}_{r}-i\tilde{B}_{i} &\tilde{B}_{\nu_{\mu}}\end{matrix}
\right)\,\,,
\end{eqnarray}
where the $B$'s are now defined in terms of the antineutrino
phase-space densities $\tilde{f}_{\nu_{e}}$,
$\tilde{f}_{\nu_{\mu}}$, $\tilde{f}_{r}$, and $\tilde{f}_{i}$. Note
that in accordance with Eq. (\ref{mixhamilton}) the coupling
coefficient in front of the integrals has to be implemented in the
equations for neutrinos with reversed sign. For two mixing neutrino
species and their antiparticles interacting with a background medium
and with neutrino-neutrino interactions included, the generalized
Boltzmann equations in their most generic form are:
\begin{widetext}
\begin{eqnarray}
\frac{\partial f_{\nu_{e}}}{\partial
t}+\mathbf{v}\cdot\frac{\partial f_{\nu_{e}}}{\partial\mathbf{r}}+
\mathbf{\dot{p}}\cdot\frac{\partial
f_{\nu_{e}}}{\partial\mathbf{p}}&=&-f_{i}\left[\frac{2\pi
c}{L}\sin2\theta
+2\left(B_{r}+\tilde{B}_{r}\right)\right]+2f_{r}\left(B_{i}+\tilde{B}_{i}\right)+\mathcal{C}_{\nu_{e}}\nonumber\\
\frac{\partial f_{\nu_{\mu}}}{\partial
t}+\mathbf{v}\cdot\frac{\partial f_{\nu_{\mu}}}{\partial\mathbf{r}}+
\mathbf{\dot{p}}\cdot\frac{\partial
f_{\nu_{\mu}}}{\partial\mathbf{p}}&=&f_{i}\left[\frac{2\pi
c}{L}\sin2\theta
+2\left(B_{r}+\tilde{B}_{r}\right)\right]-2f_{r}\left(B_{i}+\tilde{B}_{i}\right)+\mathcal{C}_{\nu_{\mu}}\nonumber\\
\frac{\partial f_{r}}{\partial t}+\mathbf{v}\cdot\frac{\partial
f_{r}}{\partial \mathbf{r}}+\mathbf{\dot{p}} \cdot\frac{\partial
f_{r}}{\partial \mathbf{p}}&=& f_{i}\left[\frac{2\pi
c}{L}\left(A-\cos2\theta\right)+B_{\nu_{e}}-\tilde{B}_{\nu_{e}}-B_{\nu_{\mu}}+\tilde{B}_{\nu_{\mu}}\right]+
\left(f_{\nu_{e}}-f_{\nu_{\mu}}\right)\left(\tilde{B}_{i}-B_{i}\right)\nonumber\\
\frac{\partial f_{i}}{\partial t}+\mathbf{v}\cdot\frac{\partial
f_{i}}{\partial \mathbf{r}}+\mathbf{\dot{p}} \cdot\frac{\partial
f_{i}}{\partial
\mathbf{p}}&=&\left(f_{\nu_{e}}-f_{\nu_{\mu}}\right)\left[\frac{\pi
c}{L}\sin2\theta+\left(B_{r}-\tilde{B}_{r}\right)\right]-f_{r}\left[\frac{2\pi
c}{L}\left(A-\cos2\theta\right)+B_{\nu_{e}}-\tilde{B}_{\nu_{e}}-B_{\nu_{\mu}}+\tilde{B}_{\nu_{\mu}}\right]\,\,.
\label{selfgenericboltz}
\end{eqnarray}
In a very similar fashion one can write the corresponding
anti-particle analog. We need to interchange tildes and reverse the
sign for $A$ to complete our set of equations:
\begin{eqnarray}
\frac{\partial \tilde{f}_{\nu_{e}}}{\partial
t}+\mathbf{v}\cdot\frac{\partial
\tilde{f}_{\nu_{e}}}{\partial\mathbf{r}}+
\mathbf{\dot{p}}\cdot\frac{\partial
\tilde{f}_{\nu_{e}}}{\partial\mathbf{p}}&=&
-\tilde{f}_{i}\left[\frac{2\pi c}{L}\sin2\theta
+2\left(\tilde{B}_{r}+B_{r}\right)\right]+2\tilde{f}_{r}\left(\tilde{B}_{i}+B_{i}\right)+\tilde{\mathcal{C}}_{\nu_{e}}\nonumber\\
\frac{\partial \tilde{f}_{\nu_{\mu}}}{\partial
t}+\mathbf{v}\cdot\frac{\partial
\tilde{f}_{\nu_{\mu}}}{\partial\mathbf{r}}+
\mathbf{\dot{p}}\cdot\frac{\partial
\tilde{f}_{\nu_{\mu}}}{\partial\mathbf{p}}&=&
\tilde{f}_{i}\left[\frac{2\pi c}{L}\sin2\theta
+2\left(\tilde{B}_{r}+B_{r}\right)\right]-2\tilde{f}_{r}\left(\tilde{B}_{i}+B_{i}\right)+\tilde{\mathcal{C}}_{\nu_{\mu}}\nonumber\\
\frac{\partial \tilde{f}_{r}}{\partial
t}+\mathbf{v}\cdot\frac{\partial \tilde{f}_{r}}{\partial
\mathbf{r}}+\mathbf{\dot{p}} \cdot\frac{\partial
\tilde{f}_{r}}{\partial \mathbf{p}}&=& \tilde{f}_{i}\left[\frac{2\pi
c}{L}\left(-A-\cos2\theta\right)+\tilde{B}_{\nu_{e}}-B_{\nu_{e}}-\tilde{B}_{\nu_{\mu}}+B_{\nu_{\mu}}\right]+
\left(\tilde{f}_{\nu_{e}}-\tilde{f}_{\nu_{\mu}}\right)\left(B_{i}-\tilde{B}_{i}\right)\nonumber\\
\frac{\partial \tilde{f}_{i}}{\partial
t}+\mathbf{v}\cdot\frac{\partial \tilde{f}_{i}}{\partial
\mathbf{r}}+\mathbf{\dot{p}} \cdot\frac{\partial
\tilde{f}_{i}}{\partial
\mathbf{p}}&=&\left(\tilde{f}_{\nu_{e}}-\tilde{f}_{\nu_{\mu}}\right)\left[\frac{\pi
c}{L}\sin2\theta+\left(\tilde{B}_{r}-B_{r}\right)\right]-\tilde{f}_{r}\left[\frac{2\pi
c}{L}\left(-A-\cos2\theta\right)+\tilde{B}_{\nu_{e}}-B_{\nu_{e}}-\tilde{B}_{\nu_{\mu}}+B_{\nu_{\mu}}\right]\,\,.
\nonumber\\
\vspace{2mm}\label{selfantigenericboltz}
\end{eqnarray}
Neutrino and antineutrino evolution are nonlinearly coupled. The
collision terms differ for neutrinos and antineutrinos. The sign
reversal of $A$ means that, dependent on the mass hierarchy, only
neutrinos or antineutrinos execute the MSW resonance. The above
nontrivially coupled set of eight equations is entirely real-valued;
yet they contain all the quantum-mechanical oscillation
phenomenology. This is the complete set of kinetic equations that
include neutrino self-interactions for the real-valued neutrino
phase-space densities $f_{\nu_{e}}$ and $f_{\nu_{\mu}}$ and the
corresponding antineutrino phase-space densities
$\tilde{f}_{\nu_{e}}$ and $\tilde{f}_{\nu_{\mu}}$.\vspace{3mm}
\end{widetext}
\section{Cross section: $\mathbf{\nu_{e}+n \longrightarrow e^{-}+p}$}
\label{appendixcross} A convenient reference neutrino cross section
is $\sigma_o$, given by
\begin{eqnarray}
\sigma_o\,=\,\frac{4G_F^2(m_ec^2)^2}{\pi(\hbar c)^4}\simeq
\,1.705\times 10^{-44}\,\text{cm}^2\,\,,
\end{eqnarray}
where $G_F$ is the Fermi weak coupling constant ($\simeq 1.436\times
10^{-49}$ ergs cm$^{3}$). The total $\nu_{e}-n$ absorption cross
section for the reaction $\nu_{e}+n \longrightarrow e^{-}+p$ is then
given by
\begin{eqnarray}
\sigma^a_{\nu_{e}n}&\sim&\,\sigma_{0}\left(\frac{1+3g_A^2}{4}\right)\,\left(\frac{\varepsilon_{\nu_e}+\Delta_{np}}{m_ec^2}\right)^2
\nonumber\\&&\times\left[1-\left(\frac{m_ec^2}{\varepsilon_{\nu_e}+\Delta_{np}}\right)^2\right]^{1/2}\,\,
, \label{ncapture}
\end{eqnarray}
where $g_A$ is the axial--vector coupling constant ($\sim -1.23$),
and $\Delta_{np}=m_nc^2-m_pc^2=1.29332$ MeV .
\bibliography{boltzmann_neutrinos_bib}
\end{document}